\documentclass[twoside]{article}

\usepackage{svg}
\usepackage{multirow} % For cell merging
\usepackage{graphicx}
\usepackage{subcaption}
\usepackage{orcidlink}

\usepackage{PRIMEarxiv}

\usepackage[utf8]{inputenc} % allow utf-8 input
\usepackage[T1]{fontenc}    % use 8-bit T1 fonts
\usepackage{hyperref}       % hyperlinks
\usepackage{url}            % simple URL typesetting
\usepackage{booktabs}       % professional-quality tables
\usepackage{amsfonts}       % blackboard math symbols
\usepackage{nicefrac}       % compact symbols for 1/2, etc.
\usepackage{microtype}      % microtypography
\usepackage{lipsum}
\usepackage{fancyhdr}       % header
\usepackage{graphicx}       % graphics
\graphicspath{{media/}}     % organize your images and other figures under media/ folder

\title{Insert-expansions for Tool-enabled Conversational Agents}
\author{
  Andreas Göldi\orcidlink{0000-0002-0631-4567}, Roman Rietsche\orcidlink{0000-0002-6112-1709}  \\
  University of St.Gallen \\
  andreas.goeldi@unisg.ch \\
}

\begin{document}
\maketitle

\begin{abstract}
This paper delves into an advanced implementation of Chain-of-Thought-Prompting in Large Language Models, focusing on the use of tools (or "plug-ins") within the explicit reasoning paths generated by this prompting method. We find that tool-enabled conversational agents often become sidetracked, as additional context from tools like search engines or calculators diverts from original user intents. To address this, we explore a concept wherein the user becomes the tool, providing necessary details and refining their requests. Through Conversation Analysis, we characterize this interaction as insert-expansion — an intermediary conversation designed to facilitate the preferred response. We explore possibilities arising from this 'user-as-a-tool' approach in two empirical studies using direct comparison, and find benefits in the recommendation domain. 
\end{abstract}

% keywords can be removed

\section{Introduction}
\label{sec:introduction}
Human language is a means both of communication and thought \cite{Kompa.2023}. In recent years, it has become much more feasible to process written language computationally. This has been achieved by mimicry of the human brain. In so-called deep learning \cite{LeCun.2015}, multiple layers of artificial neurons are used to approximate functions \cite{Hornik.1989}, for example, text to labels. Advanced deep learning architecture \cite{Vaswani.2017}, a focus on generating text \cite{Radford.}, and scaling up models \cite{Brown.28.05.2020} has led to human-like performance on natural language tasks. Aligning language outputs with human expectations \cite{Christiano.2017} and chat capabilities make the current state-of-the-art competitive with human-like performance in tasks outside of traditional natural language processing, e.g., in test-taking \cite{Bubeck.2023}. These capabilities have been appreciated by both users communicating with models in their daily lives and researchers trying to apply and advance them \cite{Liu.2023}. Since human expectations on speech have become prominent, the paramount mimicry has shifted from biological to social exemplars. Recent developments indicate that yet another source of inspiration is about to be included, namely explicit human cognition and tool-use.

These developments have been kick-started by prompting models to think through their assigned tasks step-by-step \cite{Wei.2022}. Reducing complex into simpler tasks has been advocated at least since Cartesius \cite{Dika.2023}. By chaining simple tasks together, i.e., sequentially inputting outputs from previous steps to models, language models can solve more complex problems more reliably \cite{Wu.2022}. In this way, it is possible to call a language model multiple times until a final answer is returned\footnote{This is how langchain agents operate: \url{https://python.langchain.com/docs/modules/agents/}}. Humans can think using writing \cite{Applebee.1984}, and language models imitate reasoning by generating written text, step-by-step. Since this paradigm allows intermediate steps, the idea has arisen to insert calls to tools, such as search engines, calculators, or python functions \cite{Parisi.2022}. Thereby, the language model is simulating the behavior of a computer user, and can therefore incorporate information accessible via these tools into its final answer. In this way, tool-enabled language models depart from simple function approximators, and become so-called augmented language models \cite{Mialon.2023}, with the core human capabilities of reasoning and tool-use \cite{Oakley.1957} as the exemplar to be imitated.

With social exemplars as a main source of inspiration, chat models have been trained to mimic human speech patterns \cite{Bubeck.2023}. Now that thought is imitated, less effort needs to be allocated to approximate these surface patterns, since many of them are a result of human reasoning and planning during dialog \cite{Schegloff.2007}. This means that natural speech patterns may result as a side-effect of more closely imitated reasoning paths.

The augmented language models that emulate reasoning are still meant to provide answers, and even impressive reasoning paths will not lead to user satisfaction if they remain hidden, unresponsive, and long-winded. If an answer cannot be given after one or few tool uses, augmented language models will produce intermediate observations that are ever-more divergent from the initial query, since the next step is always prompted by, i.e., conditioned on, the output of the previous steps. In this way, they tend to become side-tracked, and the primary aim, providing a satisfactory answer to the user, may be lost.

However, humans sometimes think about what they say. They raise and fulfil or deny each others' expectations; and instead of silently thinking about a final answer, humans will oftentimes probe the interlocutor, by, e.g., checking their understanding, scoping the final answer, or enhancing its appeal. This interactive nature has been extensively studied and formally described using conversation analysis \cite{Schegloff.2007}. If an appropriate response cannot be given immediately, human speakers tend to insert a new pair of utterances into the conversation, which is supposed to bridge the remaining gap. For example, if someone wants to sell you a souvenir, you will insert a question ascertaining its price before deciding on your final answer. This pattern often relies on explicit reasoning carried out in-between dialog utterances. Augmented language models already talk to themselves and to tools. There have also been recent developments which insert intermediate steps to directly ask users to provide context or check formatting for tool inputs\footnote{\url{https://python.langchain.com/docs/modules/agents/tools/how_to/human_approval}, \url{https://python.langchain.com/docs/modules/agents/tools/integrations/human_tools}}. This is a potentially powerful mechanism if used in regular chatbot interaction, since for one, it may help to avoid side-tracking in tool-enabled conversational agents because using dialog, common ground can be more easily established between interlocutors, even if one of them is a chatbot \cite{Tolzin.2023}. Furthermore, it replicates exactly the discussed feature of human talk-in-interaction, namely that of probing interlocutors to support fulfilling or reshaping the expectations they raised in their initial main utterance.

In this paper, we will therefore discuss how insert expansions may be used in tool-enabled conversational agents, and how their impact may be studied. For this, we present a paradigm of direct comparison, as well as data from one pilot and two empirical studies based on it.

\section{Background}
Before delving into the study materials and data, it pays to assimilate a better understanding of the nature of insert expansions, and to give a short overview of the nascent field of augmented language models.

\subsection{Sequence Organisation of Dialog}
\label{sec:SeqOrg}
Insert-expansions are one of several distinct building blocks of natural dialog or talk-in-interaction that can be grouped not based on topicality but on what is being done with the utterances belonging to the different blocks \cite{Schegloff.2007}. Talk-in-interaction, such as with a conversational agent, is successful if to every raised utterance, one of four responses is given: The nominally preferred response, such as agreement to an invitation; the dispreferred response, often a rejection; a temporizing response like "I may be there"; or a blocking response, such as "I have plans already" or a counter in the sense of "What about you?". This base pair of utterances is generally known as "adjacency pair". In natural dialog, such adjacency pairs are often extended by other pairs that are inserted before, between, or after the base pair, which determines the main action of a sequence. Pre-extensions either generically draw attention or gauge interest in the action performed with the specific intended base pair. If drawing attention fails or interest is low, the intended second base pair part may never be uttered. Multiple exchanges can occur before the first part of a base pair. A special kind of pre-extension is the pre-pre-extension, e.g. "Can I ask you something?" - "Yes?", which always precedes other pre-extensions, such as "You now know something about dialogues, don't you?" - "I guess I do". Normally, while pre-extensions are used by the initiator, \textit{insert-expansions} are used by the receiver, except in multi-turn inserts. They are meant to get the conversation from the raised expectation to its fulfillment. So-called post-first insert expansions serve to recover from misunderstandings and involve acknowledgement of the repair or a restatement of the first part of the base pair. So here, we can check understanding, clarify intent, or gather information. Pre-second inserts, on the other hand, ask for information required to choose between the four options for a second base pair part, for example the time and day for an invitation. This may include scoping the response but also enhancing the appeal (or at least managing expectations). After the second pair part has been uttered, a subdialog may continue with a follow-up. Very often, this happens minimally with so-called sequence closing thirds, such as "Great". Others are more wordy, for example if receivers tack on qualifications like "Just as friends, right?". Especially if the preferred response is not given, initiators may choose to challenge the response or rework the first base pair part to try again. Other options include appending closings, such as "Bye", additional talk on the topic, a topic-shift, or the initiation of a new sequence. In natural dialog, these new sequences are often of the same type by the same initiator, such as in question-series, reciprocating the main action, or following a larger action plan for the dialog. This description of sequence organization based on Schegloff's work on Conversation Analysis \cite{Schegloff.2007} has empirical support and seems to be language-universal \cite{Kendrick.2020}.

Properties of sequence organization largely generalize to text-based chats as well \cite{Tudini.2015}, even though conversations may follow slightly different patterns, which may be investigated using digital conversation analysis \cite{Meredith.2019}. However, humanizing chatbots to allow for more natural interaction is a popular way of striving for less friction in human-chatbot interaction, and bears major benefits \cite{Schanke.2021}. One way of achieving this is to explicitly anthropomorphize features of the interaction \cite{Epley.2007}. Aligning digital with natural dialog patterns fits nicely in this tradition.

Since in most applications, the users steer conversation, adjacency pair expansions of the second speaker are of primary interest for augmented language models, especially if used as conversational agents. Insert-expansions are such expansions, and potentially reduce friction in interaction as well as divergence during reasoning. They either support the first base pair part of an adjacency pair, or aim to bring about the second. Instances relevant to text-based conversations include clarifying intent, scoping responses, and enhancing appeal.

\subsection{Augmented Language Models}
Language models can generate textual inputs to arbitrary functions, which themselves may produce other text. This output can then be used as context to generate the next step. If the tool is one of information retrieval, for example, it can produce a final, grounded answer to a user query. This is how tool-augmentation of language models functions at its most basic \cite{Parisi.2022}.

Unaugmented Large Language Models display so-called formal linguistic competence, i.e., they can handle language in itself. Where they are still lacking is in functional linguistic competence, which means that they cannot do everything humans do with language. This includes formal reasoning like logic or math, using world knowledge, situation modeling in long narratives or discourses, and being able to use communicative intent as in pragmatics or establishing common ground \cite{Mahowald.2023}. Popular early tools address these issues, and therefore include information retrieval from documents, search engines, and code interpreters including calculators \cite{Mialon.2023}. Smaller, more specialized or more easily updated models can also be used as tools \cite{Xu.2023}. Considering the modular make-up of the human brain \cite{Sporns.2016}, this indicates that tooling may not only provide a capability for imitating mental activities that are deliberative in humans, but one for imitating functional brain architecture more generally, thus reaching back to a biological exemplar again.

The main advantage of augmented language models, however, is not that they are good imitators of actual biological, social, and cognitive processes. They may be. But their usefulness extends to economically more interesting opportunities as well, namely to a deepening of automation. For example, augmented language models have been applied to chemical tasks such as drug discovery \cite{Bran.2023}.

Such opportunities incentivize the further development of augmented language models. Consequently, we can expect them to become more common. This is reinforced by the fact that the setup of tools is already being automated \cite{Schick.2023}, as is the chaining of model calls \cite{Cheng.2023}. As many tools are simply python functions or calls to well-documented APIs, the code-generating capabilities of even unaugmented language models \cite{Bubeck.2023} will likely enable future augmented language models to extend their capabilities by themselves. While these advances are spearheaded by closed-sourced state-of-the-art models, augmentation is feasible with open-source models as well \cite{Xu.2023b}, even if some suggest better performance by first generating instructions on how to use the tools with such models \cite{Yang.2023}. Such instructions or function metadata may be embedded as well to make the process more efficient \cite{Hao.2023}.

Besides automating the augmentation process, and tweaking the efficiency of augmented language models, much focus is also put into keeping models running for longer, so that they can solve more complex tasks. More long-lasting runs may be enabled by extending regular chain-of-thought prompting by plan-and-solve prompting, which can generate more structured reasoning paths that otherwise would have needed to be hard-coded \cite{Wang.2023}. For these longer-running calls, decoupling observations from reasoning may serve to lessen the impact of divergence from the user intent due to misfitting observations \cite{Xu.2023c}. However, if supervision by humans is feasible, insert expansion may add additional benefits even here.

In addition to cognitive automation, tool-enabled language models may also serve to more closely align with social exemplars. As we have seen in Section \ref{sec:SeqOrg}, humans think and talk in iterative fashion to one-another. This interspersing of mutual information gathering and thinking is not idle, it serves to achieve communicative success \cite{Schegloff.2007}. Besides the benefits of humanizing chatbots \cite{Schanke.2021}, imitating this iterative fashion by introducing insert-expansions to tool-enabled conversational agents therefore is expected to have additional inherent value in terms of improved chat outcomes.

This short overview should give an idea of why we believe it is important to study augmented language models and tool-enabled conversational agents, and provide the necessary background to understand our empirical approach.

\section{Methodology}
\label{sec:Methodology}
There have been efforts to make benchmarks available for tool-augmented language models \cite{Li.2023}. However, because insert expansions require user input, they are not readily automated. Furthermore, as insert expansions help to shape final responses, these responses will depend on the idiosyncratic user input and cannot be easily standardized. This is why we decided to use human evaluations instead of benchmarking for this paper.

For this, we used fluent English speakers from prolific\footnote{\url{prolific.com}}. I our pilot, we recruited n=10 participants (age m=31.50, SD=7.53; 60\% female); in Study 1, n=71 (age m=31.45, SD=10.05; 27\% female); and in Study 2 n=36\footnote{In Study 2, we excluded two additional participants, one because their ratings had no variance while contradicting their qualitative feedback, indicating a misunderstanding in the evaluation; and the other because one of the conversational agents unaccountably asked them for their name, which the participant reported to have put them off using it.} (age m=32.22, SD=12.83; 31\% female, 1 other). Sample sizes were based on power estimations; 80\% power for a medium effect size in a one-sample, one-tailed t-test for Study 1, and 90\% for a large effect for Study 2, which was an attempt at replication with a more suitable scenario.

We will now briefly discuss the study artifact, namely a chat interface to two differently configured conversational agents, then discuss how direct comparison of the two works using this artifact, and finally, we will discuss pilot, Study 1, and Study 2 in sequence.

\subsection{Artifact}
\label{sec:artif}
To explore whether insert expansions would have an impact on tool-enabled chatbot interaction, we created two augmented language model conversational agents, one vanilla, and one with additional user-as-a-tool tools. We are going to call them vanilla and enabled bot from now on.

The agents used the python library langchain to prompt the OpenAI model gpt-3.5-turbo-0301\footnote{\url{https://platform.openai.com/docs/models/model-endpoint-compatibility}}. We used existing tools; in Study 1, the tools could query wikipedia and do simple mathematical operations; in Study 2, an embedding-based PDF-reader was included.

The user-as-a-tool tools modified the human-as-a-tool tool\footnote{\url{https://python.langchain.com/docs/modules/agents/tools/integrations/human_tools}} from the langchain library, and steered queries via websocket to the user interface. To prompt insert expansions, we simply changed name and description according to some of the purposes of insert expansions delineated in Section \ref{sec:SeqOrg}. See Figure \ref{tab:tools} for the tool definitions.

\begin{table}[]
    \centering
    \begin{tabular}{|l|}
    \hline
         clarify\_intent.description "useful if you do not understand what the appropriate type of response would be"\\
         scope\_response.description = "useful if you need more information on the human to tailor your answer to their needs"\\
         enhance\_appeal.description = "useful if you want to make your forthcoming response more appealing"\\
    \hline
    \end{tabular}
    \caption{Prompts used for insert expansion tools in Study 1; in Study 2, only scope\_response was used.}
    \label{tab:tools}
\end{table}

\begin{figure}[ht]
    \centering
    % First plot in the upper left
    \begin{subfigure}[b]{0.48\textwidth}
        \includegraphics[width=\textwidth]{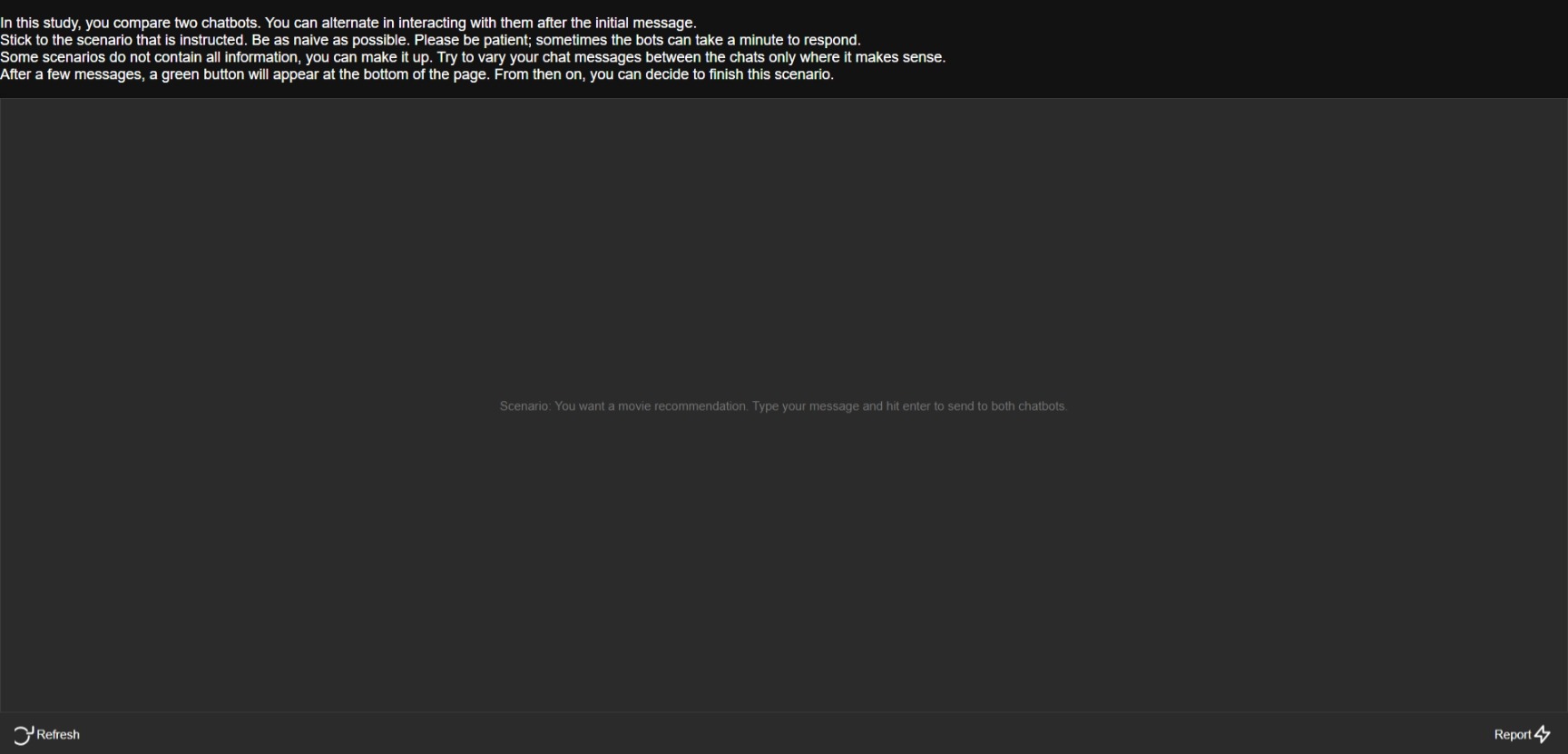}
        \phantomsubcaption % Allows for referencing without displaying the subcaption
        \label{img:scenario1}
    \end{subfigure}
    % This ensures that the next subfigure is put on the same line
    \hfill
    \begin{subfigure}[b]{0.48\textwidth}
        \includegraphics[width=\textwidth]{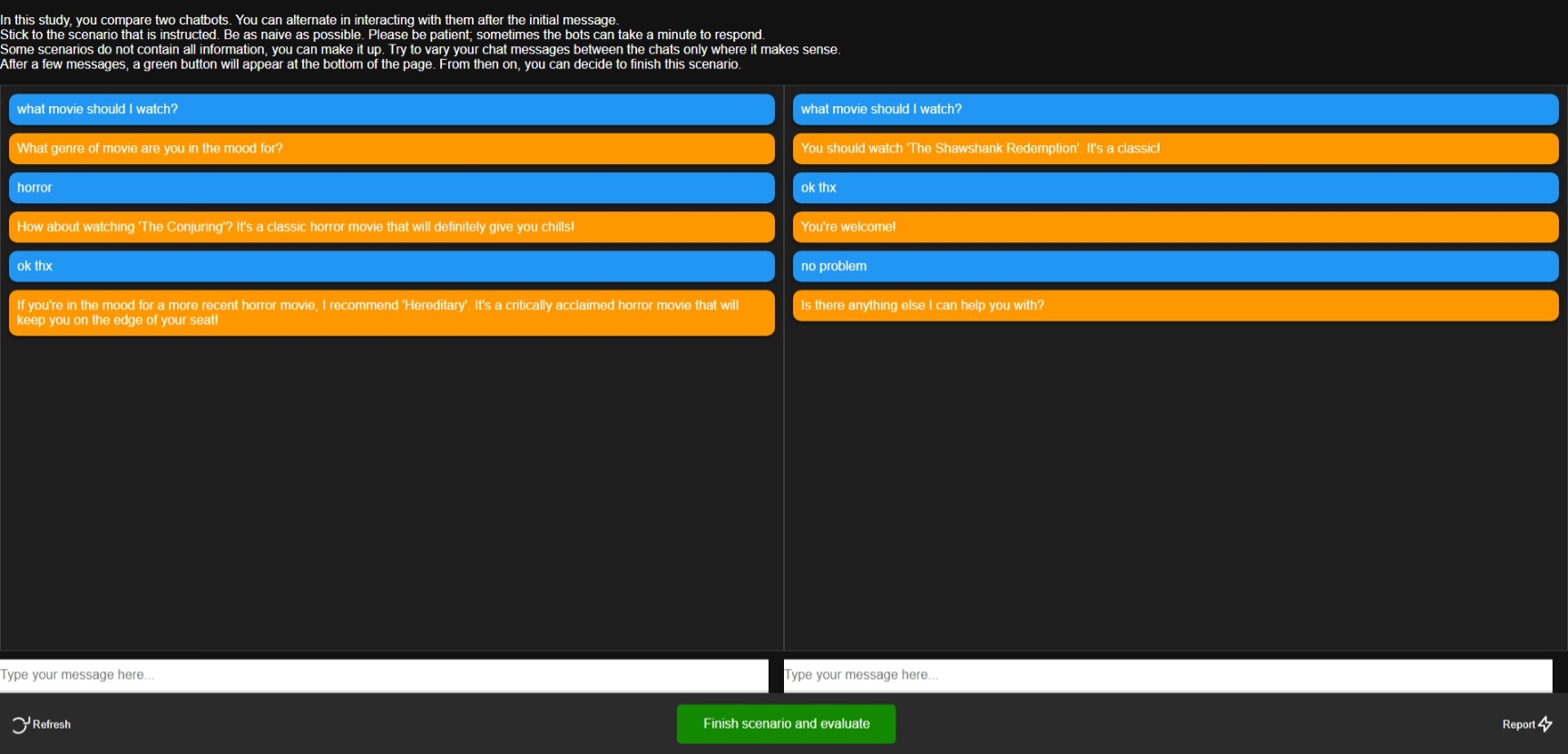}
        \phantomsubcaption
        \label{img:scenario1_f}
    \end{subfigure}
    
    % Captions and third plot in the second row
    \begin{minipage}[b]{0.48\textwidth}
        \begin{flushleft}
            Figure \ref{img:scenario1}: Before seeing the two separate chat interfaces, participants sent one query to both chatbots. \\
            Figure \ref{img:scenario1_f}: Example chat (scenario 1 is depicted). On the left side, insert expansion is enabled, the right side is a vanilla chat. This was held constant throughout the study. \\
            Figure \ref{img:rating}: Bipolar rating scales after each scenario (Study 1: 7 points; Study 2: 6 points). \\
        \end{flushleft}
    \end{minipage}
    \hfill
    % Third plot in the lower right
    \begin{subfigure}[b]{0.48\textwidth}
        \includegraphics[width=\textwidth]{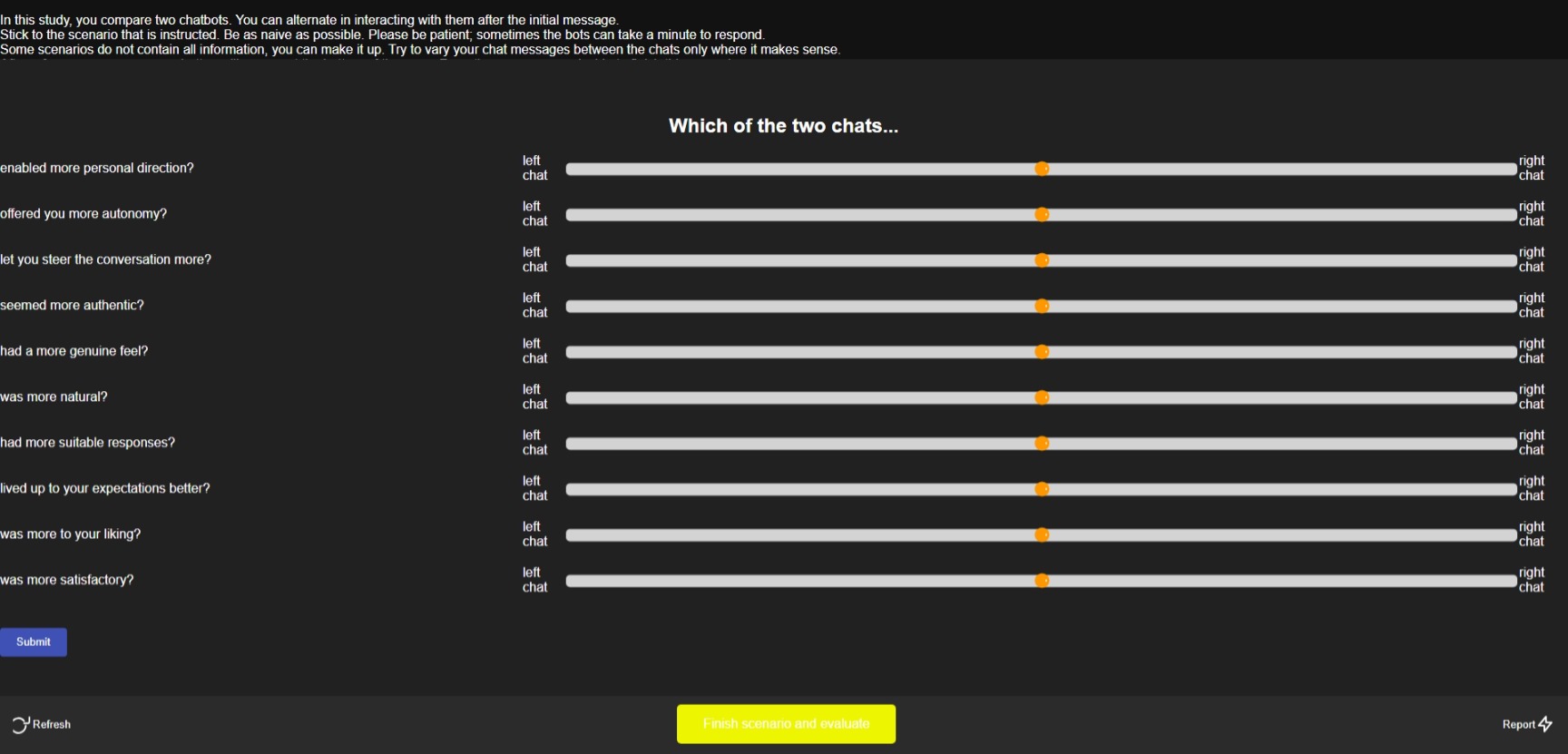}
        \phantomsubcaption
        \label{img:rating}
    \end{subfigure}
    
\end{figure}

In Figure \ref{img:scenario1}, the chat interface is shown before distinguishing between the two bots. Participants were instructed on the specific scenario via the placeholder in the input field. This was to ensure that they had to transfer the scenario to working memory before starting it. Having done so, they sent their initial query to both bots, which then appeared on each side of the screen. This stage is depicted in Figure \ref{img:scenario1_f}. They were instructed to use the two bots and compare them directly. After at least 3 bot messages per bot in Study 1 (2 in Study 2), participants could decide to finish the scenario and evaluate the bots (See evaluation modal in Figure \ref{img:rating}). 

\subsection{Direct comparison}
\label{sec:dircomp}
The aim of the studies reported herein was an initial comparison of tool-enabled conversational agents with and without insert-expansion capacity. As there was no prior literature on the specific effect, we opted to assess the differences directly, i.e., by making participants aware of both options and asking them to compare. As presenting the bots sequentially would lead to a repeated conversation, we made them available concurrently. To restrain divergence of use, we kept the scenarios short, with only a few turns. This approach allows for relatively small sample sizes and a simple study design that does not necessitate, e.g., balancing. However, any results are only of a correlational nature, as we did not randomize. We see this as appropriate for this stage of research, which is about guiding an emerging phenomenon, and not yet about establishing causal explanations.

For the comparison, we used bipolar rating scales, with the placement (left, right) of the two bots as anchors (See Figure \ref{img:rating}). We deemed this appropriate as the placement constitutes a definite reference without biasing results by naming the bots. We visually indicated the middle of the scale with a small vertical line; however, we did not label intermediary steps so as to not jeopardize equidistance \cite{Hohne.2021}. In Study 1, we allowed no-preference, in Study 2, we used a forced-choice procedure by eliminating the midpoint. The former approach does not presume a preference, and may thus be more appropriate as a measure of spontaneous preference; the latter forces participants to decide, and thus evokes cognition about preferences, which may lead to preference formation even if none was spontaneously available.  The latter approach therefore captures additional information, which may be useful, for example, if the purpose of a study is to enable managerial decision making, e.g., between two available options.

\section{Empirical Methods and Results}

\subsection{Pilot}
We used the bot usability scale 15 item version to evaluate both enabled and vanilla bots \cite{Borsci.2022}. There was not much difference between the two (m=3.59, SD=0.68 and 3.63, SD=0.25). We therefore decided not to investigate usability directly, as potential differences resulting from the inclusion of insert expansion would likely not show up in the data using our implementation.

Besides user testing the function of the bots, evaluation, and database, we also asked pilot participants to provide feedback in open answer format. This included a question on how the study was perceived, how the interface was perceived, and how the two bots differed in their estimation. There was also room to give additional feedback.

To assess the binary sentiment on the study and interface (positive or negative), we used a distilbert model \cite{Sanh.2019}\footnote{\url{https://huggingface.co/distilbert-base-uncased-finetuned-sst-2-english}}. To judge whether vanilla or enabled bots were preferred, we used a Large Language Model\footnote{gpt-3.5-turbo-0613 with system prompt "Classify if the right or the left chatbot is preferred. You can only respond with one word, 'left', 'right', 'neutral', or 'unclear', with this exact spelling." Note that this needs 2 tokens because of 'unclear'.}. Of the 10 participants, 70.00\% had positive feedback on the study, 90.00\% on the interface. Four participants preferred the enabled, two the vanilla bot (2 neutral; 2 unclear).

\subsection{Study 1}
Based on the pilot, we decided to proceed by fixing the reported bugs and replacing the usability scale with potential mediators to better assess which users would prefer the different bots. We conducted this study with n=71 participants (see Section \ref{sec:Methodology}).

\subsubsection{Measures}
In Study 1, we added 7-level bipolar rating scales for direct comparison (See Section \ref{sec:dircomp}). We were interested in how much control participants experienced; how natural they felt the chats to be; how well their intent was fulfilled in the chats; and how satisfied they were with them. See Table \ref{table:scales} for items and reliabilities of the constructs.

To account for potential mediation, we assessed interindividual difference variables, namely the most prominent personality measure, big 5, using a 15-item scale (BFI-2-XS) \cite{Soto.2017}. As insert expansions exist to smooth interaction by shaping expectations, we also assessed our participants need for cognitive closure (NFC-15) \cite{Roets.2011}.

\begin{table}[ht]
    \centering
    \begin{tabular}{|l|l|p{8cm}|}
        \hline
        \textbf{Scale} & \textbf{Items} & \textbf{Reliability} \\
        \hline
        \multicolumn{3}{|c|}{Which of the two chats...} \\
        \hline
        \multirow{3}{*}{Control} & enabled more personal direction? & \multirow{3}{=}{\textit{Study 1: $\alpha=.77$, $\lambda=.79$ (overall), for scenarios .77/.79, .84/.85, .91/.91 each; Study 2: $\alpha=.81$, $\lambda=.84$}} \\
        & offered you more autonomy? & \\
        & let you steer the conversation more? & \\
        \hline
        \multirow{3}{*}{Naturalness} & seemed more authentic? & \multirow{3}{=}{\textit{Study 1: $\alpha=.93$, $\lambda=.93$ (overall), for scenarios .93/.93, .92/.91, .97/.97 each; Study 2: $\alpha=.93$, $\lambda=.95$}} \\
        & had a more genuine feel? & \\
        & was more natural? & \\
        \hline
        \multirow{2}{*}{Intent-Effectiveness} & had more suitable responses? & \multirow{2}{=}{\textit{Study 1: $\alpha=.93$, $\lambda=.93$ (overall), for scenarios .93/.91, .95/.93, .96/.95 each; Study 2: $\alpha=.95$, $\lambda=.96$}} \\
        & lived up to your expectations better? & \\
        \hline
        \multirow{2}{*}{Satisfaction} & was more to your liking? & \multirow{2}{=}{\textit{Study 1: $\alpha=.95$, $\lambda=.95$ (overall), for scenarios .95/.95, .92/.92, .97/.97 each; Study 2: $\alpha=.98$, $\lambda=.97$}} \\
        & was more satisfactory? & \\
        \hline
    \end{tabular}
    \caption{Direct comparison bipolar rating scales with Crohnbach's $\alpha$ and Guttman's $\lambda$ 6.}
    \label{table:scales}
\end{table}

\subsubsection{Procedure}
\begin{figure}
    \centering
    \includegraphics[width=.8\linewidth, trim=0.1cm 12cm 0.1cm 12cm, clip]{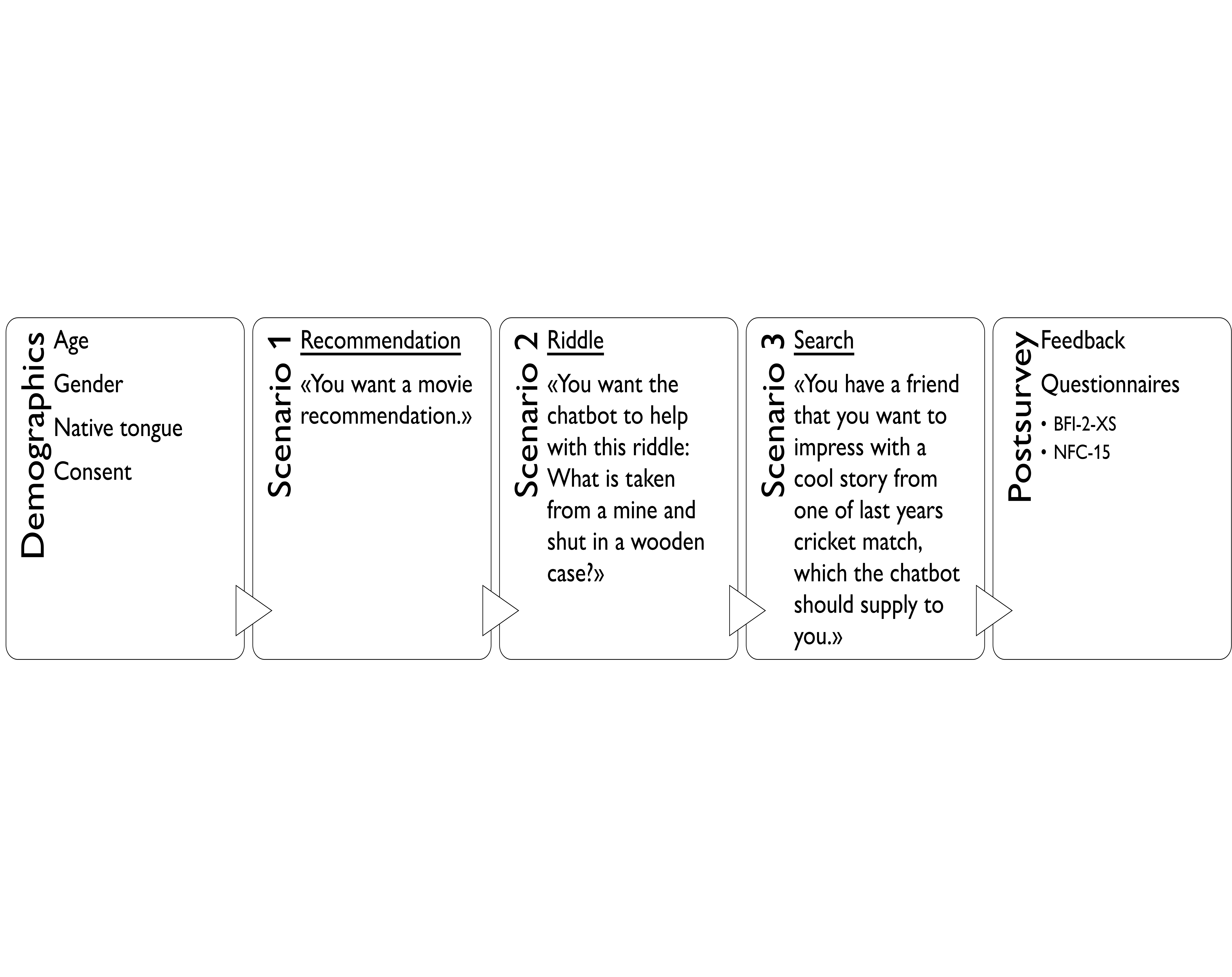}
    \caption{Study 1 procedure.}
    \label{img:procedure}
\end{figure}

After a short demographic questionnaire, n=71 participants were given 3 scenarios (see Figure \ref{img:procedure}), rating each before turning to the next (see Figure \ref{sec:dircomp}). Having completed all scenarios and evaluations, feedback was elicited and interindividual variables were assessed.

\subsubsection{Feedback}
To quantify the feedback, we repeated the procedure described for the pilot. Feedback on the study was 69.01\% positive (neutral opinions tend to be rated as negative as well, e.g. "I have no opinion about the study."), on the interface only 52.11\% (negatives include no feedback at all, neutral statements such as "It was fast and responsive, just feel like the loading is too big and the lettering also", but also some on bugs like "It was fine, although sometimes my prompt would trigger a loading animation that the bot would never reply to, so I had to prompt again, which left the loading anim on the screen for one of the bots but not the other, not a big deal." or "It was frusting when it could not listen or answer all my question"). 46.48\% of participants preferred the enabled bot (16.90\% preferred vanilla; 2.82\% neutral, and the rest unclear). While there was a tendency to see the enabled bot [i.e, the left bot] as more personable, opinions differed, for example, one participant stated "The chatbot on the left had a more personal touch to it. The one on the right felt more like a robot and more focused on facts." but another "The one on the left was robot-like and the one on the right was very human". This goes for overall preference too: "Not sure but I felt the first one was more helpful." versus "I think the right one was more intelligent, I mean, it gave better responses". Other participants gave feedback that nicely puts the two options into contrast, such as "Both are good, but I think the left one is more cautious, and the one on the right answers faster.", "The chat on the left was a lot quicker and asked more questions in terms of my desired answer. The chat on the right was more strict to the point.". So, in summary, adding insert-expansion capability does not automatically make conversational agents superior, especially as the vanilla bot is already trained to fulfill social expectations. Even if this imitation is only on the surface, it is apparently enough for many users. % these statements could be mined for design principles, dont include too much here; there are also some more severe bug reports in there, altough most have been taken care of already
\subsubsection{Results}
See Table \ref{tab:descr} for descriptive statistics, Table \ref{table:scales} for reliabilities of the direct comparison and Section \ref{sec:dircomp} for the methodology. The measures are not constant across situations, which means retest-reliability could not be assessed. Intra-class correlation (ICC) using average random raters were ICC=.63 for control, ICC=.61 for naturalness, ICC=.34 for intent-effectiveness, and ICC=.47 for satisfaction. Crohnbach's $\alpha$ for a one-factor structure was .96.

\begin{table}[ht]
\centering
\label{table:summary_statistics}
\begin{tabular}{|lrrrrrrr|}
  \hline
Variable & Mean & SD & Median & Min & Max & Skew & Kurtosis \\ 
  \hline
Age & 31.45 & 10.06 & 29.00 & 18.00 & 69.00 & 1.39 & 2.01 \\ 
Gender* & 1.73 & 0.45 & 2.00 & 1.00 & 2.00 & -1.03 & -0.96 \\ 
O & 3.76 & 0.79 & 4.00 & 1.33 & 5.00 & -0.43 & 0.04 \\ 
C & 3.75 & 0.87 & 3.67 & 1.67 & 5.00 & -0.47 & -0.40 \\ 
E & 2.72 & 0.89 & 2.67 & 1.33 & 4.67 & 0.56 & -0.23 \\ 
A & 3.78 & 0.71 & 4.00 & 2.00 & 5.00 & -0.60 & -0.40 \\ 
N & 2.82 & 1.16 & 2.67 & 1.00 & 5.00 & 0.02 & -0.93 \\ 
NFC & 3.88 & 0.67 & 3.90 & 2.13 & 5.47 & -0.15 & 0.34 \\ 
Control Situation 1 & -0.32 & 1.45 & -0.33 & -3.00 & 3.00 & 0.04 & -0.50 \\ 
Naturalness Situation 1 & -0.23 & 1.73 & 0.00 & -3.00 & 3.00 & 0.09 & -0.88 \\ 
Intent Effectiveness Situation 1 & -0.27 & 1.83 & 0.00 & -3.00 & 3.00 & 0.15 & -1.11 \\ 
Satisfaction Situation 1 & -0.35 & 1.85 & -1.00 & -3.00 & 3.00 & 0.27 & -0.95 \\ 
Control Situation 2 & -0.11 & 1.54 & -0.17 & -3.00 & 3.00 & 0.05 & -0.76 \\ 
Naturalness Situation 2 & 0.24 & 1.79 & 0.33 & -3.00 & 3.00 & -0.25 & -0.98 \\ 
Intent Effectiveness Situation 2 & 0.61 & 1.91 & 1.00 & -3.00 & 3.00 & -0.63 & -0.73 \\ 
Satisfaction Situation 2 & 0.34 & 2.00 & 1.00 & -3.00 & 3.00 & -0.39 & -1.12 \\ 
Control Situation 3 & -0.22 & 1.65 & -0.33 & -3.00 & 3.00 & 0.31 & -0.63 \\ 
Naturalness Situation 3 & -0.11 & 1.82 & -0.17 & -3.00 & 3.00 & 0.22 & -0.86 \\ 
Intent Effectiveness Situation 3 & 0.29 & 1.94 & 0.25 & -3.00 & 3.00 & -0.30 & -1.01 \\ 
Satisfaction Situation 3 & 0.08 & 1.99 & 0.00 & -3.00 & 3.00 & -0.08 & -1.13 \\ 
   \hline
\end{tabular}
\caption{Summary statistics of all measures in Study 1. Negative values for the situational measures indicate that insert expansion provided more control, naturalness, intent-effectiveness, or satisfaction.}
\label{tab:descr}
\end{table}

\begin{figure}
    \centering
    \begin{minipage}{.48\textwidth}
        \centering
        \includegraphics[width=.8\linewidth]{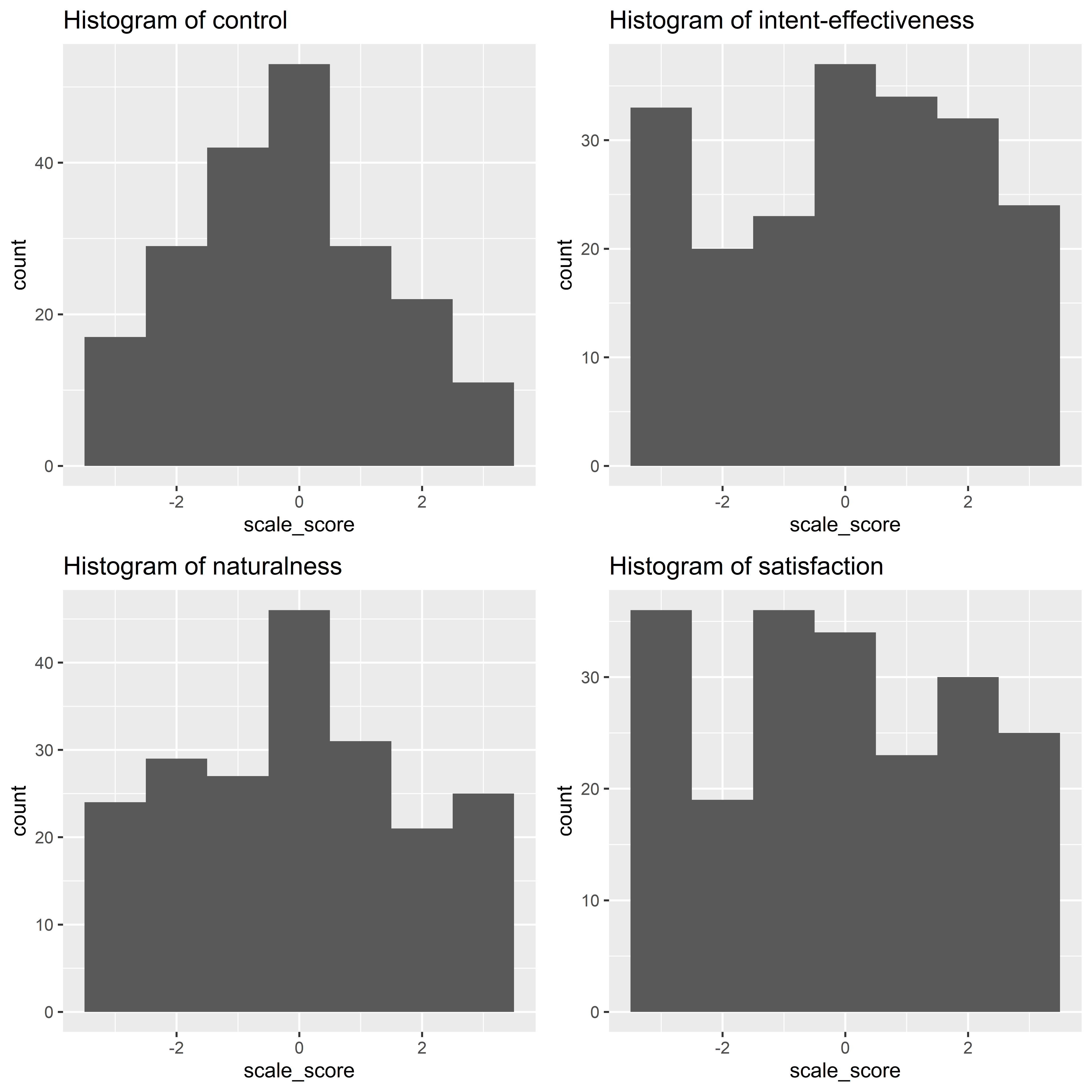}
        \caption{Distribution of ratings in Study 1.}
        \label{fig:histsstudy1}
    \end{minipage}%
    \begin{minipage}{.48\textwidth}
        \centering
        \includegraphics[width=.8\linewidth]{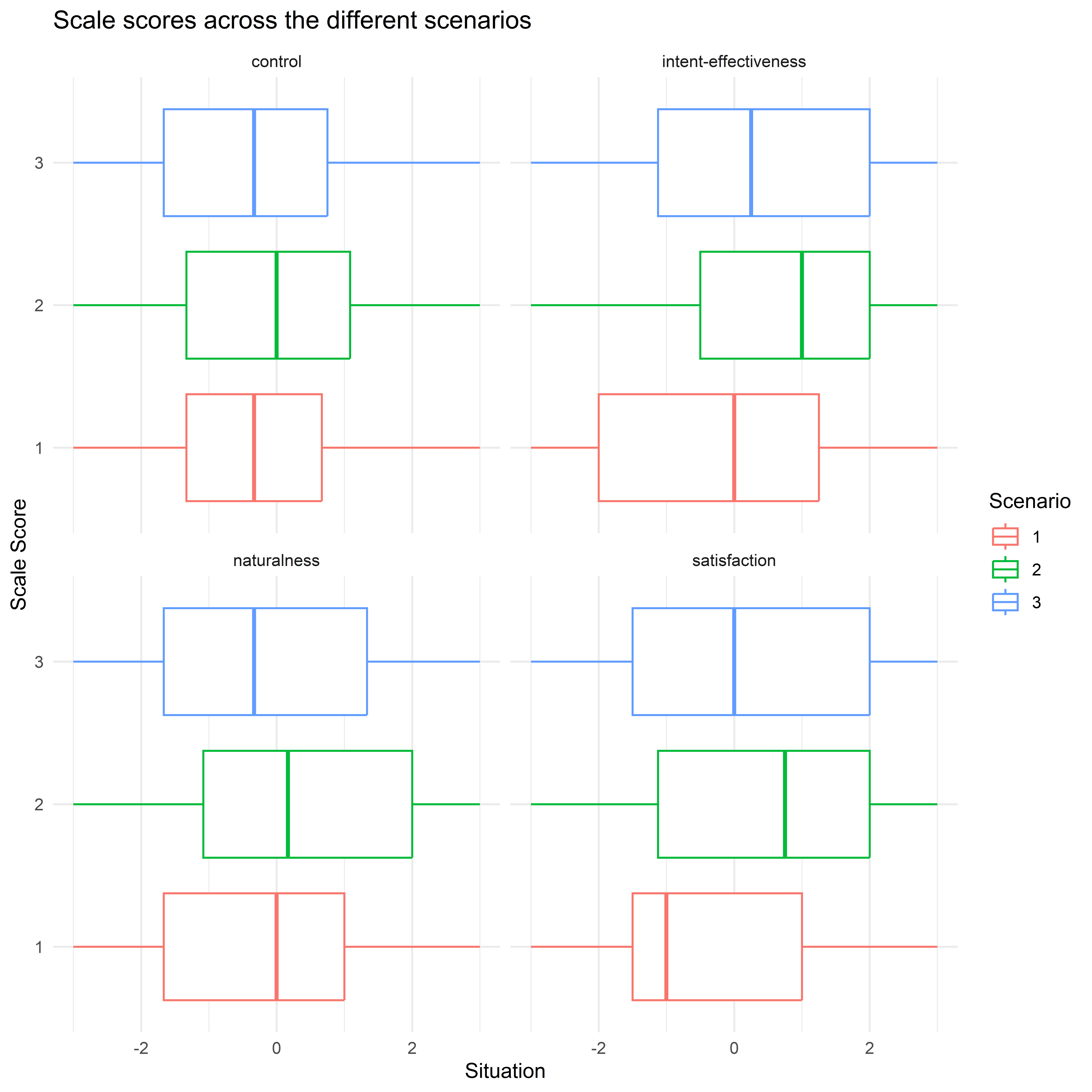}
        \caption{Direct comparison results in Study 1 (negative values indicate preference for enabled, positive for vanilla bots).}
        \label{fig:scalescores}
    \end{minipage}
\end{figure}

The distribution of control ratings was normal; of naturalness normal with noticeable but statistically acceptable kurtosis (-0.88); of intent-effectiveness seemingly bimodal, with a population at the extreme end of the enabled bot, however, with the main distribution at an offset towards the vanilla bot; and satisfaction almost uniform, or potentially multimodal as well, meaning normality assumptions were not fulfilled here; see Figure \ref{fig:histsstudy1}.

The interindividual variables all approximated normal distributions, with expected ranges. Notably, openness, agreeableness, and conscientiousness were relatively high (out of 5) with 3.76 (SD=.79) 3.75 (SD=.87) 3.78 (SD=.71), cutting off the extreme positive tail.

In terms of significance, only control in scenario 1 differed between the two bots (the difference in satisfaction for scenario 1 was marginal with p=0.05985, as tested with a Wilcoxon Signed Rank test). Control in scenario 1 is normally distributed; and a difference was observable in a one-sample, one-tailed t-test (t=-1.8494, df=70, p=0.03431). In the other situations, control was not significantly different from 0, although the tendency persists. This is because the effect size seems to be small (for this t-test, Cohen's d is d=.22). In the other two scenarios, no significant differences were observed (see Figure \ref{fig:scalescores}). Overall preference (treating all scales as one factor) is marginally not significant with V=827, p=0.05463 in a Wilcoxon Signed Rank test.

\begin{figure}
    \centering
    \includegraphics[width=.9\linewidth]{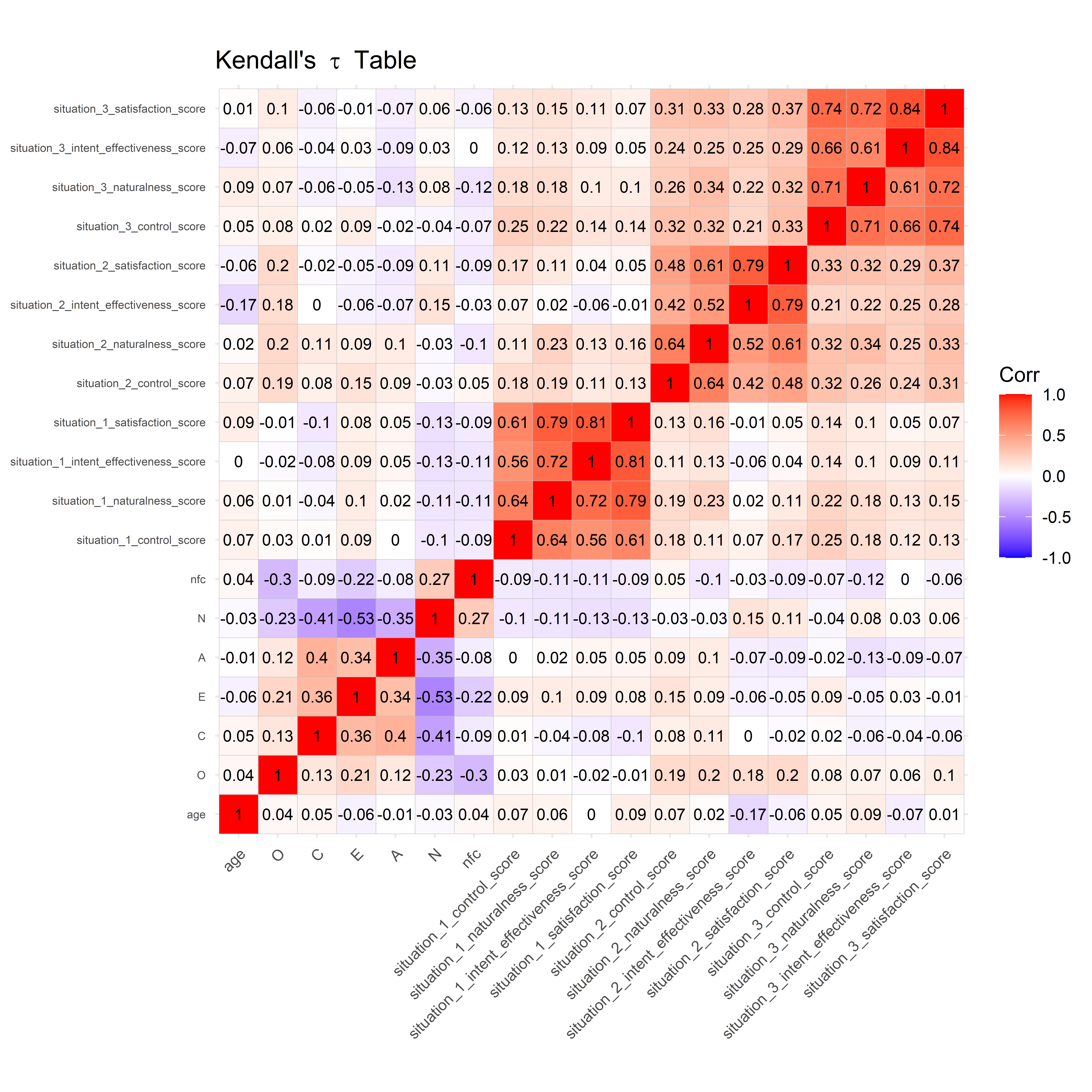}
    \caption{Kendall's $\tau$ plot for all variables in Study 1.}
    \label{fig:tau}
\end{figure}

Looking at the associations between variables, and especially potential mediators, Table \ref{fig:tau} provides an overview. There is visible clustering in the scenarios for the direct comparison measures. For scenario 2, where the bots were supposed to help with solving a riddle, openness seemed to be important across ratings. However, most associations do not exceed $\tau$=.2.

\subsubsection{Discussion}
In summary, the impact of adding insert-expansions to tool-enabled conversational agents on user experience depends on the specific use case of the agent. The described measures varied between scenarios, with ICCs under .5 for intent-effectiveness and satisfaction. There was more consistency for control and naturalness, which should therefore be put into more focus for generalizable statements about the role of adding insert-expansions. 

In this study, we only observed one significant difference, namely for control in scenario 1 (a recommendation scenario). However, taking into account the open answer feedback, this may have been the result of heterogeneous preferences. More specifically, of bimodal distributions, especially in intent-effectiveness, and potentially in satisfaction, meaning that preference depends either on individual user characteristics or specific breakdowns or successes in the particular chats. This is plausible based on the role of insert expansions in conversational repair and the gentle shaping of expectations \cite{Schegloff.2007}. Satisfaction may even become observably trimodal in larger samples, as some seem to prefer the enabled bot, some the vanilla bot, while others do not have any preference. This may be especially likely if some conversations failed, be it due to excessive or counterproductive insert expansions, or the lack of insert expansions where they would have been helpful.

\subsection{Study 2}
As the necessity of adding insert-expansions depends on scenarios, we added a new scenario for a follow-up study. We chose: "You want to work on the most important sustainable development goal in the 2022 UN report but do not know which it is. Type your message and hit enter to send to both chatbots." In this scenario, both bots had an information retrieval tool and access to the United Nations annual report of 2022 (which is after the training data of the underlying model ended). They were prompted to ask for a grounded recommendation, as Study 1 indicated recommendation as a relevant domain to introduce insert-expansion capabilities. We reduced the number of insert-expansion prompts for the user-as-a-tool tool in the enabled bot, only retaining the scope\_response tool (see Section \ref{sec:artif}). In Table \ref{tab:example}, the difference between chats with and without insert expansion are illustrated.

We conducted this study with n=36 participants (see Section\ref{sec:Methodology}).

\begin{table}
    \centering
    \small
    \begin{tabular}{|c|c|}
        \hline
        \textbf{Vanilla} &
        \textbf{User-as-a-tool} \\
        \hline
            \multicolumn{2}{|c|}{\textbf{Query:} What is the most important sustainable development goal in the UN annual report 2022 I could work on?} \\
        \hline
        \begin{minipage}{0.55\textwidth}
            \textit{Entering new chain...} \\
            I should check the UN annual report for 2022 to find the most important sustainable development goal. \\
            \textbf{Action}: UN info \\
            \textbf{Action Input}: Search for the UN annual report for 2022 \\
            \textbf{Observation}: I don't know the answer. \\
            \textbf{Thought}: I should try searching for the specific sustainable development goals mentioned in the report. \\
            \textbf{Action}: UN info \\
            \textbf{Action Input}: Search for sustainable development goals in the UN annual report for 2022 \\
            \textbf{Observation}: The annual report for 2022 mentions the sustainable development goals several times. It highlights the support provided to 162 countries and territories to rescue the goals, the delivery of \$17 billion in operational activities for development, the facilitation of \$95 million in trade investment deals through support to South-South agreements, the protection of 138 million workers from work-related deaths, injuries, and disease, the prevention of 77 million tons of CO2 emissions through clean energy initiatives, the support provided to 183 million children with access to remote learning, and the convening of a conference on sustainable transport as a means to achieve the sustainable development goals. \\
            \textbf{Thought}: Based on the information provided in the UN annual report for 2022, it is difficult to determine the most important sustainable development goal. However, some of the goals mentioned in the report include rescuing the goals, delivering operational activities for development, supporting South-South agreements, protecting workers, preventing CO2 emissions, providing access to remote learning, and promoting sustainable transport. It would be helpful to further analyze the report to identify the specific goals that are highlighted as the most important. \\
            \textbf{Final Answer}: The most important sustainable development goal in the UN annual report for 2022 is not specified. Further analysis of the report is needed to determine the most important goal. \\
            \textit{Finished chain.} \\
        \end{minipage} & 
        
        \begin{minipage}{0.35\textwidth}
            \textit{Entering new chain...} \\
            I need to find the UN annual report for 2022 to determine the most important sustainable development goal. \\
            \textbf{Action}: UN info \\
            \textbf{Action Input}: UN annual report 2022 \\
            \textbf{Observation}: I don't know the answer to the question as it is not provided in the given context. \\
            \textbf{Thought}: I need to use the scope\_response tool to gather more information about the person's specific interests or field of work. \\
            \textbf{Action}: scope\_response \\
            \textsc{First pair part of insert expansion (bot query):}\\
            \textbf{Action Input}: What is your field of work or area of interest? \\
            \textsc{Second pair part of insert expansion (user input):}\\
            \textbf{Observation}: Finance \\
            \textbf{Thought}: I can now provide a tailored answer based on the person's interest in finance. \\
            \textbf{Final Answer}: The most important sustainable development goal in the UN annual report 2022 that you could work on is Goal 8: Decent Work and Economic Growth. This goal focuses on promoting sustained, inclusive, and sustainable economic growth, full and productive employment, and decent work for all. As someone interested in finance, you could contribute to this goal by supporting initiatives that promote job creation, entrepreneurship, and financial inclusion. \\
            \textit{Finished chain.} \\
        \end{minipage} \\
        \hline 
    \end{tabular}
    \caption{Two example reasoning paths that have access to a pdf reading tool and the 2022 UN report on sustainable development goals: \url{https://www.un.org/annualreport/files/2022/09/SG-Annual-Report-2022_E_1.pdf}. The example illustrates how 'user-as-a-tool' can help to reign in tool-enabled language models, and provide more useful answers. Note that a user only sees the final answer (and the query "What is your field of work or area of interest?" in the user-as-a-tool version).}
    \label{tab:example}
\end{table}

\subsubsection{Measures}
We retained all measures from Study 1. However, we exchanged the 7-level bipolar rating with midpoint for a 6-level forced-choice bipolar rating. In this, the midpoint was still visually indicated; however, participants had to decide whether they would rather indicate preference for the enabled or vanilla bot. We used the forced-choice approach because in Study 1, many ratings tended towards the midpoint, meaning that only some participants spontaneously formed a preference; however, formed preferences is what we were interested in, and therefore, we elicited preference formation via forced choice. 

\subsubsection{Procedure}
\begin{figure}
    \centering
    \includegraphics[width=.6\linewidth, trim=0cm 10cm 0cm 0cm,clip]{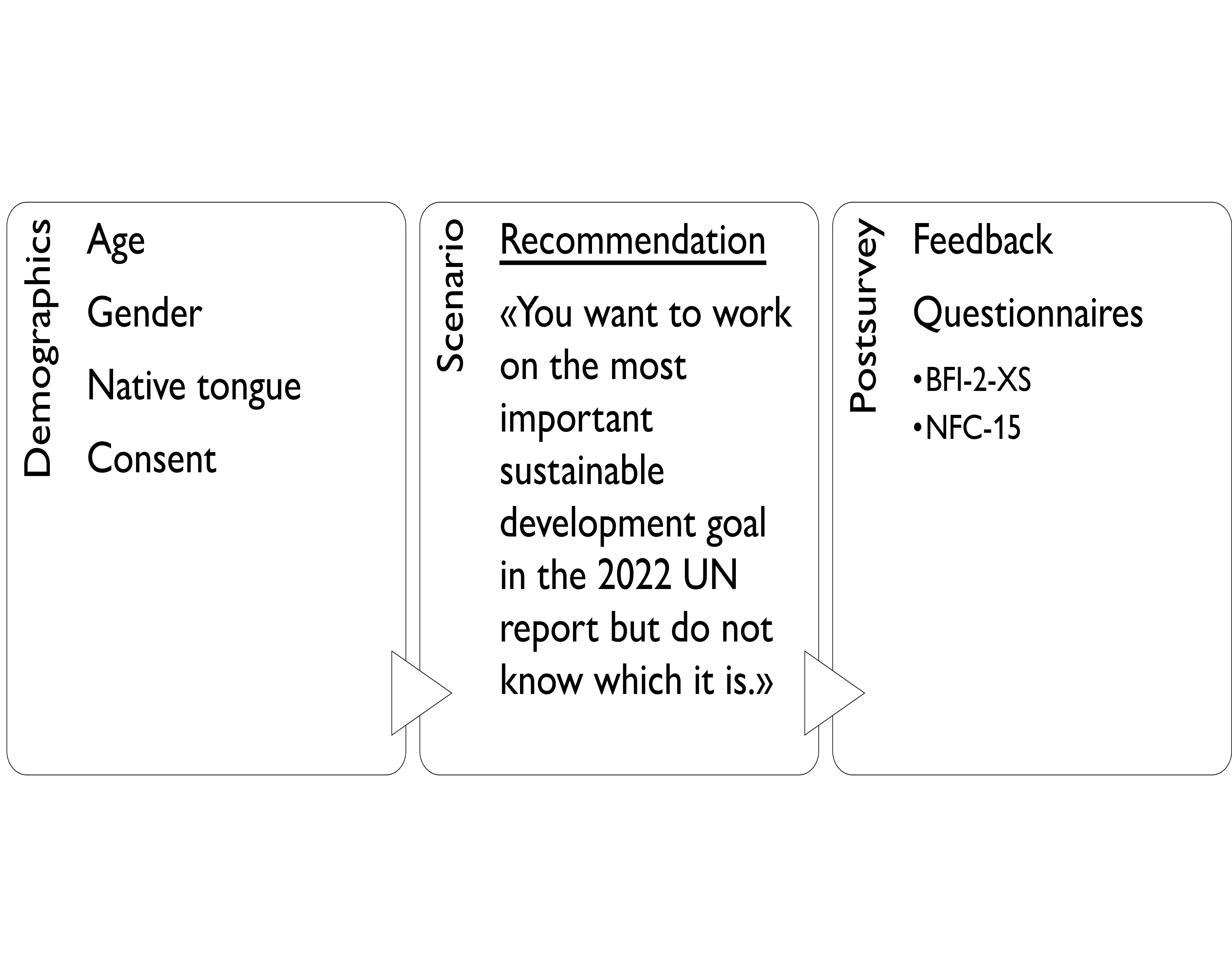}
    \caption{Study 2 procedure.}
    \label{img:procedure2}
\end{figure}

The procedure remained the same as in Study 1, with the exception that only one scenario was given (for more details, see Figure \ref{img:procedure2}).

\subsubsection{Feedback}
Both study and interface feedback was positive with 66.67\% each. 36.11\% preferred the enabled version, 27.77\% vanilla, while 13.88\% were neutral, and the rest of the preferences unclear. Qualitatively, the feedback largely conformed with the one to Study 1.

\subsubsection{Results}
See Table \ref{tab:descr2} for descriptive statistics, and refer for reliabilities again to Table \ref{table:scales} and for the methodology to Section \ref{sec:dircomp}. One-factor internal consistency was $\alpha$=.95.

\begin{figure}
    \centering
    \begin{minipage}{.48\textwidth}
        \centering
        \includegraphics[width=.8\linewidth]{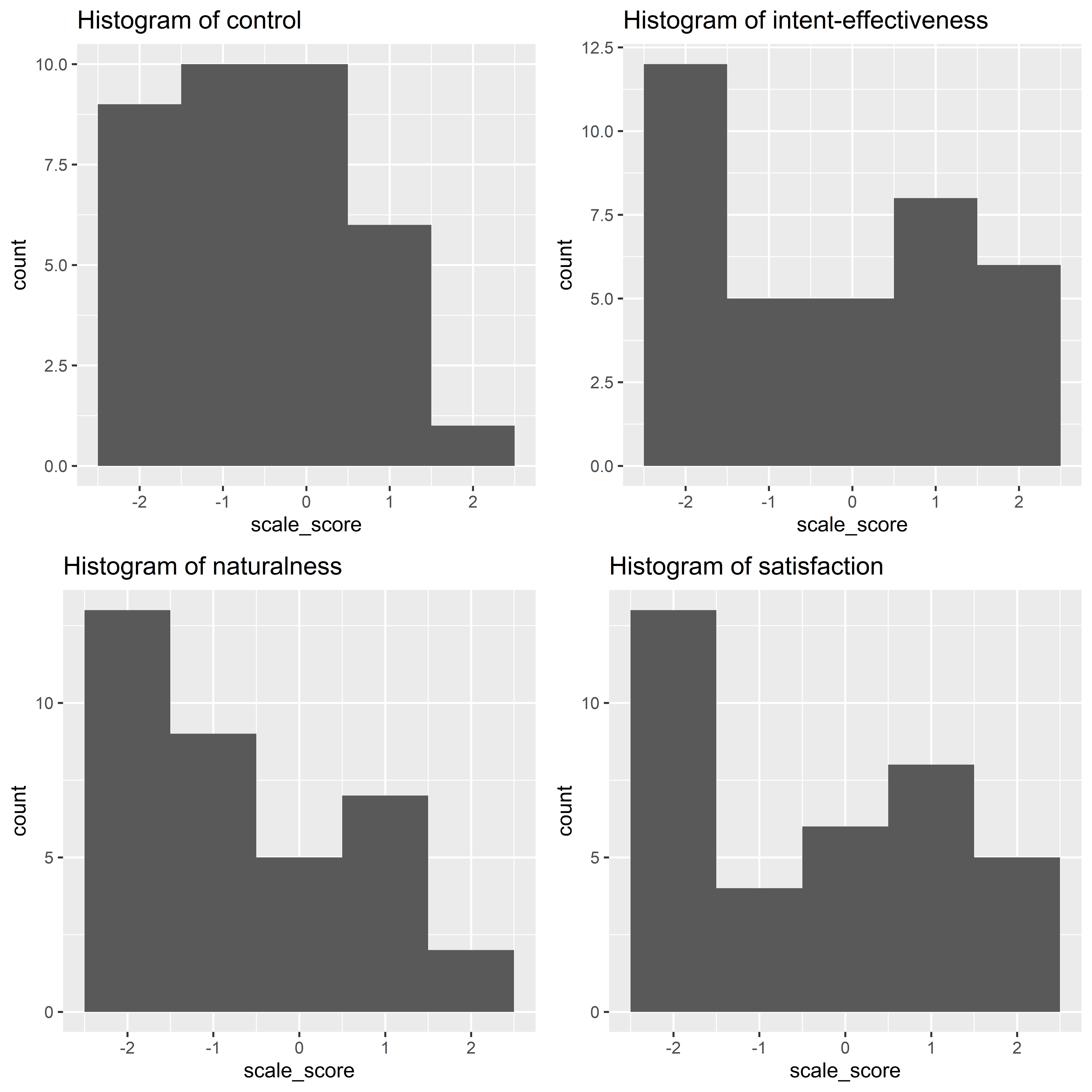}
        \caption{Distribution of ratings in Study 2.}
        \label{fig:histsstudy2}
    \end{minipage}%
    \begin{minipage}{.48\textwidth}
        \centering
        \includegraphics[width=.8\linewidth]{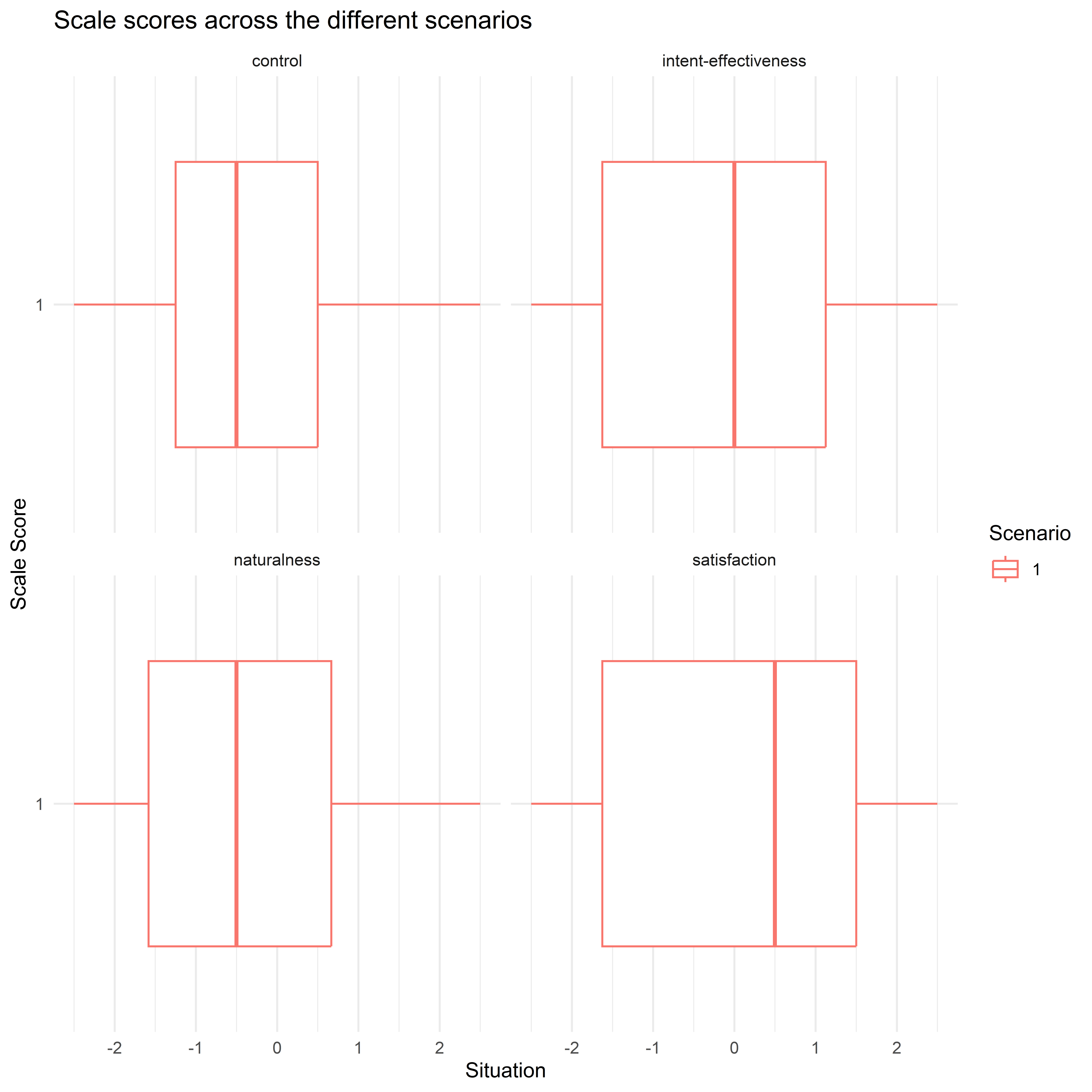}
        \caption{Direct comparison results in Study 2 (negative values indicate preference for enabled, positive for vanilla bot).}
        \label{fig:scalescores2}
    \end{minipage}
\end{figure}

For distributions, see Figure \ref{fig:histsstudy2}. The smaller sample size leads to less clean normal distributions, rendering non-parametric tests safer. We can still observe potential bimodality in intent-effectiveness and satisfaction, specifically regarding clumping at the extreme end of preference for the enabled bot, which is out of the normal distribution. Extreme preference for the vanilla bot may have been reduced by the choice of scenario.

In this study, we observed significant differences from null preference in control and naturalness, both towards the enabled bot (using Wilcoxon Signed Rank tests, for control: V=217, p=0.03427; for naturalness: V=208, p=0.02484). For intent-effectiveness and satisfaction, effects were non-significant (V=271, p=0.3266 and V=299, p=0.2976, respectively), with a tendency towards the vanilla bot (see Figure \ref{fig:scalescores2}). Overall preference (treating all scales as one factor) is, as in Study 1, not significantly different from null preference (V=216, p=0.08296).

\begin{table}[ht]
\centering
\label{table:summary_statistics_reduced}
\begin{tabular}{|lrrrrrrr|}
  \hline
Variable & Mean & SD & Median & Min & Max & Skew & Kurtosis \\
  \hline
Age & 32.22 & 12.83 & 26.50 & 18.00 & 67.00 & 1.07 & 0.03 \\
Gender* & 1.72 & 0.51 & 2.00 & 1.00 & 3.00 & -0.32 & -0.68 \\
O & 3.87 & 0.80 & 4.00 & 2.00 & 5.00 & -0.62 & 0.23 \\
C & 3.62 & 0.78 & 3.67 & 2.00 & 5.00 & -0.24 & -0.57 \\
E & 2.91 & 0.86 & 3.00 & 1.00 & 4.67 & -0.32 & -0.36 \\
A & 3.81 & 0.70 & 4.00 & 2.00 & 4.67 & -0.85 & -0.26 \\
N & 2.59 & 1.29 & 2.33 & 1.00 & 5.00 & 0.35 & -1.28 \\
NFC & 3.74 & 0.43 & 3.77 & 2.67 & 4.60 & -0.24 & -0.36 \\
Control Situation 1 & -0.42 & 1.23 & -0.50 & -2.50 & 2.50 & -0.01 & -0.71 \\
Naturalness Situation 1 & -0.50 & 1.49 & -0.50 & -2.50 & 2.50 & 0.32 & -1.03 \\
Intent Effectiveness Situation 1 & -0.12 & 1.80 & 0.00 & -2.50 & 2.50 & 0.03 & -1.44 \\
Satisfaction Situation 1 & -0.14 & 1.80 & 0.50 & -2.50 & 2.50 & -0.00 & -1.51 \\
  \hline
\end{tabular}
\caption{Summary statistics of all measures in Study 2. Negative values for the situational measures indicate that insert expansion provided more control, naturalness, intent-effectiveness, or satisfaction.}
\label{tab:descr2}
\end{table}

\begin{figure}
    \centering
    \includegraphics[width=.7\linewidth]{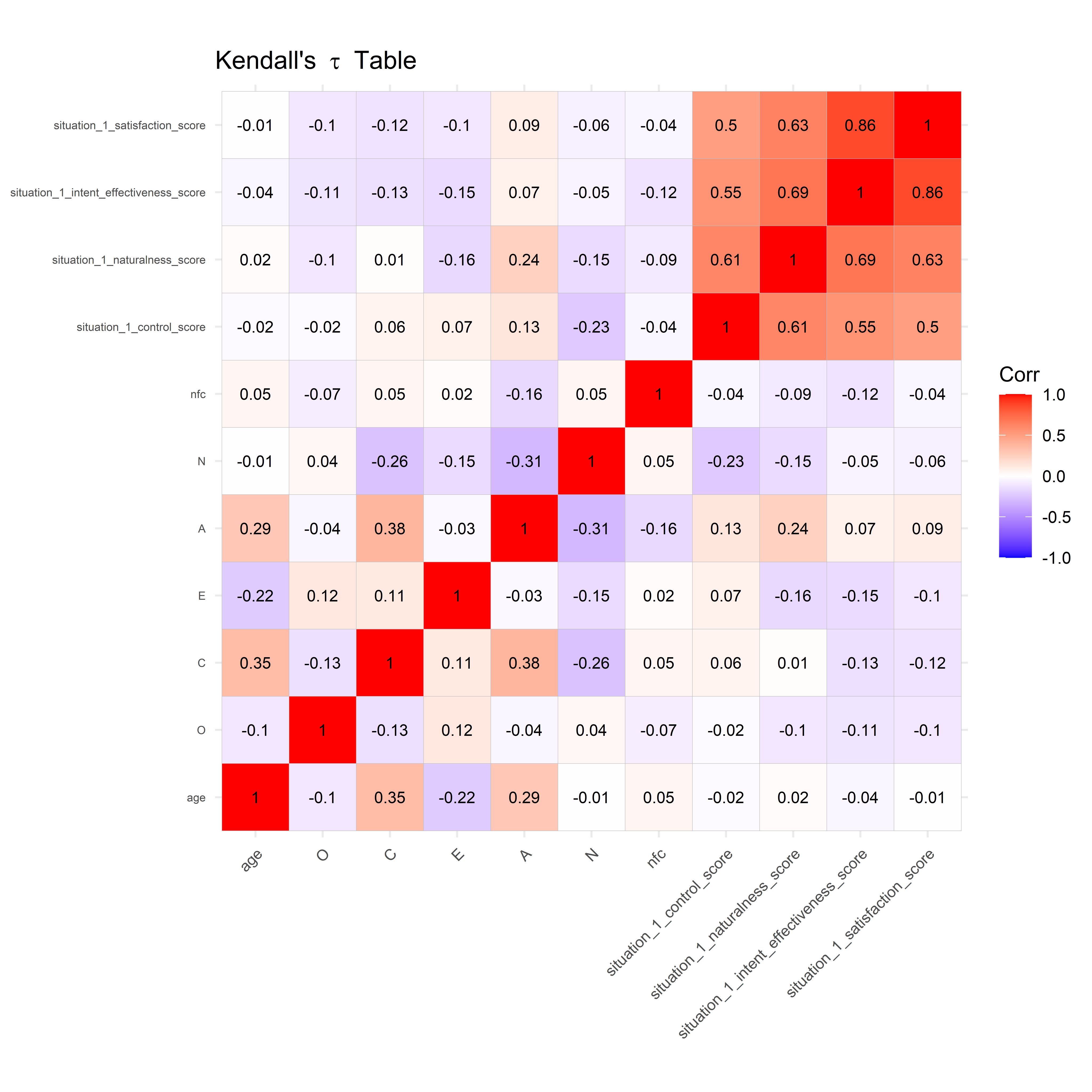}
    \caption{Kendall's $\tau$ plot for all variables in Study 2.}
    \label{fig:tau2}
\end{figure}

The variables were interrelated in a similar fashion as in Study 1 (See \ref{fig:tau2}). The largest association between situational and interindividual variables was $\tau$=.24 between agreeableness and naturalness (however, non-significantly, with z=1.8626, p=0.06252).

\subsubsection{Discussion}
Study 2 replicated a preference for insert-expansion in tool-enabled conversational agents in the recommendation domain. It also extended the marginal difference from null preference in naturalness from Study 1, with a significant difference in the modified scenario.

Heterogeneous preferences remain an observable fact in this study as well. However, the assessed mediators were associated with only small effect sizes, similar to Study 1. It was not fruitful to incorporate the assessed interindividual variables into more complex models of participants' preference, as the small magnitude of associations would have required larger samples to produce useful results.

\section{General Discussion}
In this study, we introduced the notion of 'user-as-a-tool'. This notion refers to enabling tool-enabled conversational agents with the capacity to use insert-expansions in their conversations, e.g., uttering follow-up questions or enhancing the appeal of their final answer before formulating it. This way of conversing corresponds more closely to human dialog, where many utterances only extend the main interaction \cite{Schegloff.2007}. Insert expansions are inserted in-between an initial first part of this main action, and the second part, which in tool-enabled conversational agents corresponds to the final response. Recent developments involving tool-enabled or augmented language models beckon a trend towards more imitation of cognitive exemplars in the application of artificial intelligence models \cite{Parisi.2022}, including those underlying conversational agents. The more tools augment language models, and the more expansive these tool-capabilities become, the more likely will language models imitate long-stretched reasoning paths and look-ups. The longer the reasoning path, the more detached will answers become from initial requests. That means that, especially for conversational agents, enabling tools may introduce detrimental consequences, in both longer experienced latencies by users, who cannot see the reasoning a model imitates, and potentially worse answers, since the prompting chain slowly diverges from user input. 

To explore whether such detriments could be addressed by introducing insert expansions to tool-enabled conversational agents, we carried out two studies and a pilot. In these studies, we found that insert-expansions may be useful only in certain scenarios or use cases. We believe this may be because the underlying model of both vanilla and insert-expansion-enabled conversational agents already conformed excellently to social expectations, since it has been trained to do so. In this way, even though social exemplars are imitated in the vanilla version only on a surface level, meaning that follow-up questions or scoping arise without prior reasoning, this difference in sources is often not distinguishable to users.

However, especially within appropriate use cases, such as recommendation, mimicry of natural dialog organization may be inferior to imitating dialog that includes reasoning and insert-expansion. For a grounded recommendation use case, we replicated a preference for insert-expansion in terms of control, and provide evidence to suggest a similar effect in terms of naturalness as well.

Our results indicate that this difference in the experience of the chats is heavily dependent on the specific scenario. We could not observe the expected effects for help in a riddle with incomplete information and search. This may be due to several reasons. For one, the use cases may not be appropriate. Potentially, the specific scenarios operationalizing these use cases were not implemented in a fitting fashion. Another possibility is that such scenarios would require a longer conversation for effects to materialize. Furthermore, even though our main outcome measures show excellent reliability, their validity cannot be ascertained numerically, since we did not include measures of convergent and discriminant validity.

The studies reported herein rely on direct comparison in one-sample cases. Our results, therefore, cannot be treated as causal explanations. Rather, they serve to establish knowledge about the differences and associations of variables relevant to adding insert-expansion capabilities to tool-enabled conversational agents. 

Yet another limitation of this study is that we did observe bimodality in intent-effectiveness and satisfaction in both studies, however, without providing a clear explanation of how the two populations differ. One of them seems to follow a normal distribution with an offset slightly towards the vanilla agent that did not have insert-expansion capability, and the other clumped at the very extreme of preferring the insert-expansion enabled agent. Based on the assessed potential mediators, big 5 personality and need for closure, this difference is unlikely to be entirely due to interindividual factors. If so, we did not assess the particular factor. More likely is the explanation that this second population which preferred the added insert-expansion capability was the minority directly benefiting from this capability in terms of the quality of answers or absence of conversational breakdown. We have already seen that for most use cases, surface-level mimicry of human dialog patterns may be sufficient. Only in cases where mimicry does not suffice will there be genuine and observable differences in intent-effectiveness and satisfaction with conversational agents.

This observation is analogous to the initial motivation for the studies reported herein. Augmented language models imitate not only talk, but also thought. This double use of language competence leads them to outperform unaugmented versions in many instances. However, reasoning comes with a drawback, namely that of diverging from the original user-provided prompt due to ever-more influence of the context added through reasoning and tool-use. It also takes longer to arrive at an answer, which means that user experience suffers if responses only achieve parity with the unaugmented model. By introducing insert-expansion capabilities, we can return much control and disciplining power to the user, rendering it less likely that augmented language models veer off path. However, follow-up questions need to be answered, which takes more effort than simply awaiting a response. And if the follow-up questions are badly chosen, or do not lead to any new relevant information, they may even lead the final answer astray.

So, to conclude, adding insert-expansion comes with drawbacks too. Conversations may be longer and require more effort from the user. Exactly like tool-enabled conversational agents, insert-expansion-enabled agents can only serve to improve conversations if they achieve more than parity in their final answers. This they do in recommendation scenarios, at least in terms of perceived user control. They are likely to do so in other use cases as well, and it pays to investigate them. The only caveat is that the use of insert-expansion, similar to the use of tool-enabling, should be constrained to cases where they are superior to the best-guess-quick-fire approach of existing vanilla models, which have been excellently trained to fulfill social expectations.

Closer imitation of human talk-in-interaction, based on cognitive besides social exemplars, is only beneficial to the performance of conversational agents if surface-level mimicry does not suffice to satisfy users. However, in scenarios that do necessitate the inclusion of additional information, reasoning and tools will provide benefits, and if this information can only come from the user, the user-as-a-tool approach of adding insert-expansion capabilities to conversational agents will do so as well.

\section*{Acknowledgments}
This study was supported as part of the Innosuisse Flagship SCESC.

%Bibliography
\bibliographystyle{unsrt}  
\bibliography{references}

\end{document}